\PassOptionsToPackage{numbers,sort&compress,square}{natbib}

\documentclass[floatfix,%
 aip,
 amsmath,amssymb,
 reprint,%
]{revtex4-1}

\usepackage{graphicx}
\usepackage{dcolumn}
\usepackage{bm}

\usepackage[utf8]{inputenc}
\usepackage[T1]{fontenc}
\usepackage{mathptmx}
\usepackage{etoolbox}
\usepackage{siunitx}
\usepackage[normalem]{ulem}
\usepackage{tikz}

\usepackage[numbers,sort&compress,square]{natbib}
\setcitestyle{super,sort&compress}

\usepackage{booktabs} 
\usepackage[normalem]{ulem}
\usepackage[none]{hyphenat}
\makeatletter
\def\@email#1#2{%
 \endgroup
 \patchcmd{\titleblock@produce}
  {\frontmatter@RRAPformat}
  {\frontmatter@RRAPformat{\produce@RRAP{*#1\href{mailto:#2}{#2}}}\frontmatter@RRAPformat}
  {}{}
}%
\makeatother
\begin{document}

\preprint{AIP/123-QED}

\title[]{Neutral Barium in Solid Neon: Optical Spectroscopy and First Excited State Lifetime}

\author{Alessandro Lippi *}
\affiliation{Istituto Nazionale di Fisica Nucleare, Sezione di Ferrara, 44122 Ferrara, Italy}
\affiliation{Department of Physics and Earth Science, University of Ferrara, 44122 Ferrara, Italy}
\email{lpplsn@unife.it}
\author{Giovanni Carugno}
\affiliation{Istituto Nazionale di Fisica Nucleare, Sezione di Padova, 35131 Padua, Italy}
\affiliation{National Quantum Science and Technology Institute, 00185 Rome, Italy}

\author{Roberto Calabrese}
\affiliation{Istituto Nazionale di Fisica Nucleare, Sezione di Ferrara, 44122 Ferrara, Italy}
\affiliation{Department of Physics and Earth Science, University of Ferrara, 44122 Ferrara, Italy}

\author{Federico Chiossi}
\affiliation{Istituto Nazionale di Fisica Nucleare, Sezione di Padova, 35131 Padua, Italy}
\affiliation{Department of Physics and Astronomy, University of Padua, 35131 Padua, Italy}

\author{Marco Guarise}
\affiliation{Istituto Nazionale di Fisica Nucleare, Sezione di Ferrara, 44122 Ferrara, Italy}
\affiliation{Department of Physics and Earth Science, University of Ferrara, 44122 Ferrara, Italy}

\author{Madiha M. Makhdoom}
\affiliation{Istituto Nazionale di Fisica Nucleare, Sezione di Padova, 35131 Padua, Italy}
\affiliation{Department of Physics and Astronomy, University of Padua, 35131 Padua, Italy}

\author{Giuseppe Messineo}
\affiliation{Istituto Nazionale di Fisica Nucleare, Sezione di Padova, 35131 Padua, Italy}
\affiliation{National Quantum Science and Technology Institute, 00185 Rome, Italy}

\author{Jacopo Pazzini}
\affiliation{Istituto Nazionale di Fisica Nucleare, Sezione di Padova, 35131 Padua, Italy}
\affiliation{Department of Physics and Astronomy, University of Padua, 35131 Padua, Italy}

\date{\today}

\begin{abstract}

Matrix isolation spectroscopy enables probing atomic properties in controlled cryogenic environments. 
Here we present a spectroscopic study on neutral barium atoms embedded in a neon cryogenic crystal at 6.8 K, extending previous investigations performed in other noble gas hosts. 
\\
The visible and near-infrared emission spectra were recorded under two different laser excitation schemes.
First, 10-ns laser pulses at 355 nm were used to directly excite high-lying energy levels of barium, enabling the observation of fluorescence cascades.
Second, a tunable continuous-wave laser operating between 700 nm and 900 nm allowed us to determine the matrix-induced shifts of barium energy levels relative to their vacuum values, as well as the inhomogeneous linewidths of the observed transitions and to perform lifetime measurements.
\\
Our results confirm multiple radiative pathways and matrix-induced relaxation channels affecting the 5d6s and 6s6p barium manifolds.
Furthermore, we present the first lifetime measurement of the barium 5d6s;$^3$D$_1$ state in a neon crystal, yielding 0.39 $\pm$ 0.02 s, with a predicted increase of about 10\% at 2 K.
\\
The study of fluorescence and spectroscopic properties of barium isolated in neon represents an important step toward future searches for the electron electric dipole moment using barium monofluoride in neon matrices, where neutral barium atoms may act as unavoidable impurities and potential sources of background and systematic limitations.

\end{abstract}

\maketitle

\section{\label{sec:level1} Introduction}

In recent years, spectroscopy of atoms and molecules embedded into noble gas cryogenic crystals has developed significantly, as it plays a key role for high precision measurements within the matrix isolation technique (MIT). MIT enables the confinement of highly reactive species (dopants) within an inert, low‑temperature host \cite{DUBEY2018317}, to investigate their characteristics and behavior while interrogating them at a fixed point in space and simultaneously achieving a high density of dopants in the crystal \cite{PhysRevLett.125.043601}.
\\
Other techniques to investigate atoms or molecules include slowing a production beam \cite{Patterson2007} or trapping them through Magneto Optical Traps (MOTs) \cite{Raab1987}. These methods enable in-gas investigation of the species, at the cost of the number of molecules. However, for fundamental physics research, having a large number of atoms or molecules enhances the sensitivity of the measurements \cite{Safronova2018}. Therefore, in this study, we discuss the production of an inert cryogenic environment with a high density of barium (Ba) atoms as dopants to investigate them through spectroscopic measurements as a testbench for future studies of more complex molecules such as barium monofluoride (BaF).
\\
The investigation method used is Laser Induced Fluorescence (LIF) as it allows for high signal to noise ratio and good sensitivity for atomic or molecular spectroscopy \cite{Kinsey1977}.
\\
Neutral barium has a favorable electronic structure to be observed in noble gas matrices \cite{Callear1973, Trenary1973}, and has been recently characterized in matrices such as Ar, Kr and Xe using LIF spectroscopy, revealing multiple occupancy sites and Jahn-Teller splitting \cite{McCaffrey, McCaffrey_BaSr_inRGmatrices}. Further studies in solid helium have also provide new insights on how barium, under very low perturbation conditions, shows a very long lifetime of atomic metastable states of the orders of seconds \cite{Weis_BaHelium}.
\begin{figure*}[t]
    \centering
    \includegraphics[width=\textwidth]{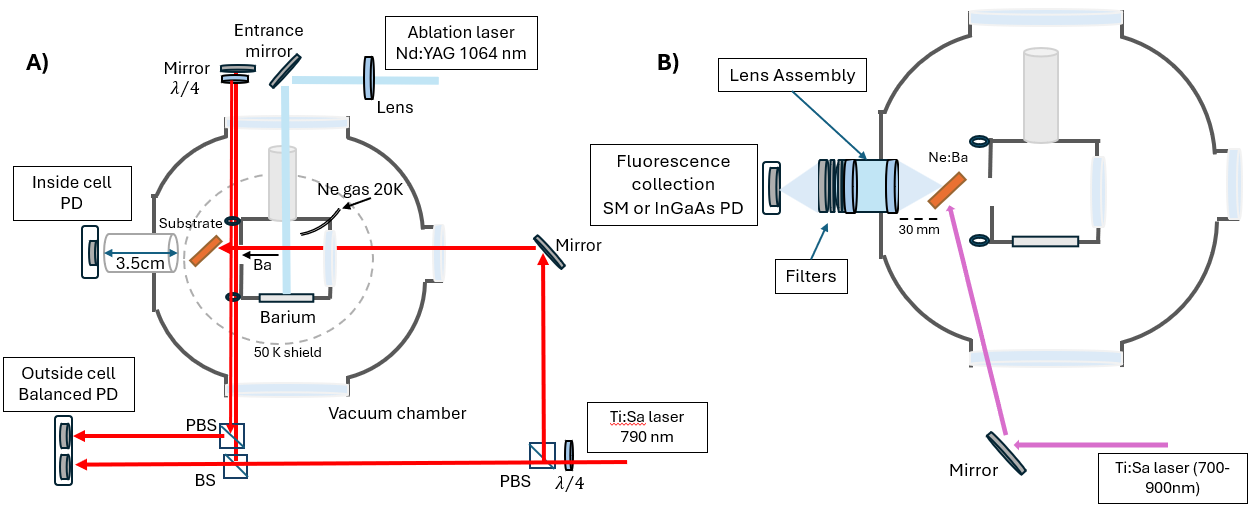}
    \caption{Top view sketch of the experimental setup including production of a gas phase barium beam, crystal growth, excitation, and fluorescence detection: A) Ba doped neon crystal growth setup: two lasers are simultaneously used. An Nd:YAG laser (light blue arrows in the figure) is used for the ablation of the metallic barium target, which gets extracted from the cell through a hole in front of a CaF$_2$ substrate. An Avesta laser (red arrows in the figure) at 12636 cm$^{-1}$ (791.13 nm) is used to ensure efficient production both inside and outside the cell during ablation. 
    B) Excitation and fluorescence detection scheme: the gas composed of Ba and neon is sprayed onto the CaF$_2$ substrate (orange rectangle) at 6.8 K, which is permanently tilted to minimize the collection of excitation light (purple arrows) in the fluorescence detection system. A custom built lens assembly allows us to fix the first collection lens at about 30 mm from the crystal growth spot, enabling efficient collection of the emitted signal which is then focused on either a spectrometer or a photodiode. }
    
    \label{fig:setupSS}
\end{figure*}
In this paper, we extend the approach conducted in previously cited works, proving the first successful growth of good quality Ba doped neon crystals at temperatures ranging from 6.8 to 9 K. Among the hosts previously studied, neon is a promising matrix for MIT, as it can offer good optical transparency and reduced perturbation of embedded species, as proven in previous studies of atoms embedded in cryogenic neon crystals \cite{1Yb_Neon_Lambo, dargyte2021optical_Rb_Neon_Spin, braggio2022spectroscopy_RbNEon}. Therefore, this approach enables direct comparison of Ba properties in our neon crystal lattice with respect to the previously studied ones. 
\\
In this work, neon serves both as a buffer gas in our atomic source and as a matrix deposition gas for crystal growth. Our setup permits LIF spectroscopic measurements of barium atoms in the crystal, using single and double laser excitation in either continuous or pulsed mode. These methods enable selective excitation of barium energy levels. We also measure, for the first time, the lifetime of the metastable state 5d6s $^3$D$_1$ of barium in a neon matrix.
\\
The study of barium in cryogenic inert matrices serves as an initial spectroscopic approach towards molecules, such as BaF,  with great characteristics for low energy, high precision measurements such as the electron electric dipole moment (eEDM) through the MIT.
\\
BaF production will include deposition of some neutral barium presence in the matrix, unless it is filtered out \cite{yau2024specularreflectionpolarmolecules, corriveau2024matrixisolatedbariummonofluoride} at the cost of a loss in the number of molecules. Therefore, understanding the spectral signatures and dynamics of Ba will be essential, as it will provide a crucial spectroscopic baseline for future BaF matrix isolation measurements in a cryogenic inert environment such as neon \cite{Li_2023, corriveau2024matrixisolatedbariummonofluoride} or parahydrogen \cite{Messineo_pH, Borghesani_2023}, as recently proposed by our group within the DOCET experiment.

\section{Experimental setup}

The experimental procedure of our spectroscopic work consists of three main stages: the growth of a neon cryogenic crystal with embedded Ba atoms and LIF detection outside the vacuum chamber.

\subsection{Barium doped neon crystals production}
\begin{figure}[]
    \centering   
    \includegraphics[width=75mm]{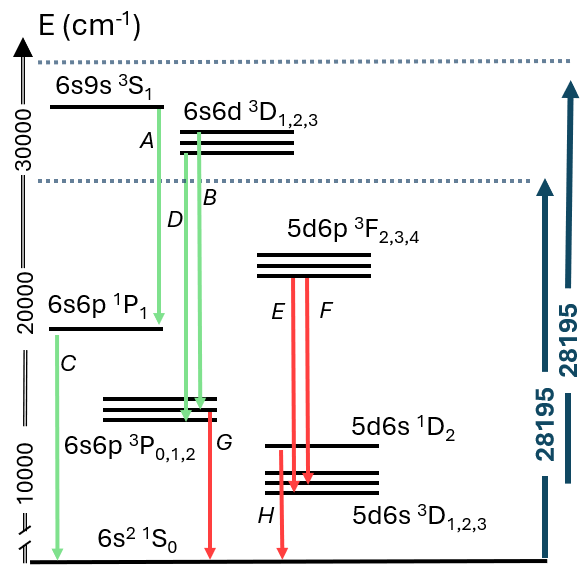}
    \caption{ Energy level diagram (not to scale)  of neutral barium showing the electronic states relevant to this work. The ground state 6s$^2\; ^1\text{S}_0$  is shown alongside the low-lying excited states 5d6s $^1\text{D}_2$ and 5d6s $^3\text{D}_{1,2,3}$. Higher-lying manifolds include $6s6p \; ^1\text{P}_1$, 6s6p $ ^3\text{P}_{0,1,2}$, 5d6p $ ^3\text{F}_{2,3,4}$, 6s6d $^3\text{D}_{1,2,3}$ and 6s9s $^3$S$_1$. These states are involved in the excitation and fluorescence processes discussed throughout the paper. Red and green lines indicates fluorescence cascade transitions identified and shown in Table \ref{Table_3rd_harm}. Blue lines represent third harmonic excitation at 355 nm. Labels A to H next to the atomic transitions are the same as the one reported in Table \ref{Table_3rd_harm}. } 
    \label{all_energy_liness}
\end{figure}

Neutral barium atoms are produced in a custom designed stainless steel cryogenic ablation cell, place in a vacuum chamber maintained below $1\times10^{-7}$ mBar as shown in Figure \ref{fig:setupSS} A), exploiting the buffer gas cooling technique \cite{doyle1995buffer, bethlem2002cryogenicbufgas}. Laser ablation, within this method, allows for the production of large numbers of atoms at very low temperature, as collisions with the cold buffer gas cool them down. In our setup, neon is used as a buffer gas. The cell is cooled with a pulse tube refrigerator Leybold-Heraeus RG210 and kept at approximately 20 K, to avoid neon condensation while still allowing efficient cooling of the produced atoms. 
\\
Ablation laser pulses at 1064 nm with a repetition rate of 10 Hz and energies between 10 and 30 mJ, generated from a EKSPLA NL300 Nd:YAG laser, hit the metallic barium target, fixed on one side of the cell, on a focused spot with diameter of approximately 0.5 mm. The plasma plume evaporates Ba atoms, carried out of the cell by neon flowing at 5–20 SCCM through a 3 mm aperture. The ablation spot is periodically shifted to maintain stable production. 
\\
To probe the production of neutral barium atoms in the gas
phase, before deposition, two absorption measurements are
performed simultaneously, as already done in previous studies
for BaF \cite{mooij2024novel}, inside and outside of the cell. An Avesta Ti:Sa laser is used, with a linewidth of approximately 2 GHz, tunable over a 14\,300 - 11100 cm$^{-1}$ range (700 - 900 nm). The laser is tuned to 12\,636 cm$^{-1}$ (791.13 nm), probing the transition from the ground state 6s$^2$ $^1\text{S}_0$ to 6s6p $3\text{P}_1$. The barium energy levels used in this work are shown in Figure \ref{all_energy_liness}. An absorption measurement probes the production directly inside the cell, where the laser is sent through a window at the back of the cell and exits from the extraction hole towards a Thorlabs DET36A/M silicon photodiode. The second method consist in a differential measurement with a Thorlabs PDB210A/M. To do so, a laser beam is equally split, one part is used as reference signal, while the other passes 5 mm from the extraction hole, in a dual pass configuration, aligned through two diaphragms at the side of the extraction hole. Noise is highly suppressed in this configuration, allowing for the detection of bunches of atoms coming out of the source every ablation pulse. 
\\
Crystal growth happens on a CaF$_2$ substrate mounted on a copper holder that is connected to a two-stage helium pulse tube cryocooler Sumitomo RP62B, maintaining a temperature of 6.8 K. A heater is positioned on top of the cold finger, allowing gradual temperature ramps and, around 16 K, full sublimation of the neon crystal and its dopants. 
\\
Following the production of Ba atoms, the neon buffer gas carries them to the CaF$_2$ substrate, where the crystal grows for 20 to 60 minutes. Before initiating the ablation process, a 10 SCCM neon gas flow is maintained for 10 minutes, allowing optimal base growth of a pure neon crystal and ensuring better sublimation of the doped crystal.

\subsection{Optical setup for fluorescence collection}
Fluorescence measurements are performed using multiple laser systems, hitting the doped crystal positioned at 45 degrees with respect to the incoming barium beam, as shown in Figure \ref{fig:setupSS} B). A combination of two identical Avesta Ti:Sa laser is used to deliver 10–200 mW on a 1 mm wide spot size on the crystal surface for single and double laser excitation, as described in Section \ref{avesta}. Pulses of 28\;195 cm$^{-1}$ (354 nm) are generated through the third harmonic of a 1064 nm Nd:YAG laser. At a repetition of 10 Hz, the pulses are 10 ns long and on a range between 1 to 2.5 mJ and are used to excite neutral barium to high energy states, as describe in the next section.
\\
Light collection occurs almost orthogonally from the excitation beam, as illustrated in Figure \ref{fig:setupSS} in purple arrows, to reduce the unwanted excitation signal and optimize the fluorescence detection.
\\
The light detection system consisted of a two lens assembly to collect fluorescence emitted from the substrate. The lenses are mounted inside a custom-built cylindrical holder that extended into the cryogenic chamber, allowing us to set the first lens at a distance of approximately 30 mm from the crystal surface, as shown in Figure \ref{fig:setupSS}. To isolate the fluorescence signal from the excitation laser light, spectral filters are used, depending on the measurement done. Detection of the fluorescence light occurred through an Hamamatsu InGaAs photodiode for lifetime measurements, or two spectrometers: Spectral Products SM642 for the visible range and Ocean Optics NIR Quest InGaAs for near-infrared (NIR) measurements of the first excited state of barium.

\section{Results and Discussion}

In this section, we present the spectroscopic characterization of neutral barium atoms isolated in a neon matrix at 6.8 K. We first analyze the high-energy excitation regime using pulsed third harmonic (355 nm) laser light, followed by the investigation of low lying transitions probed through narrow linewidth continuous wave lasers in the 700 - 900 nm range. Finally, we report the lifetime measurement of the metastable 5d6s $^3$D$_1$ state in this environment.

\subsection{High energy excitation of barium in neon matrix}
\label{Barium_third}
\begin{figure}[]
    \centering   
    \includegraphics[width=90mm]{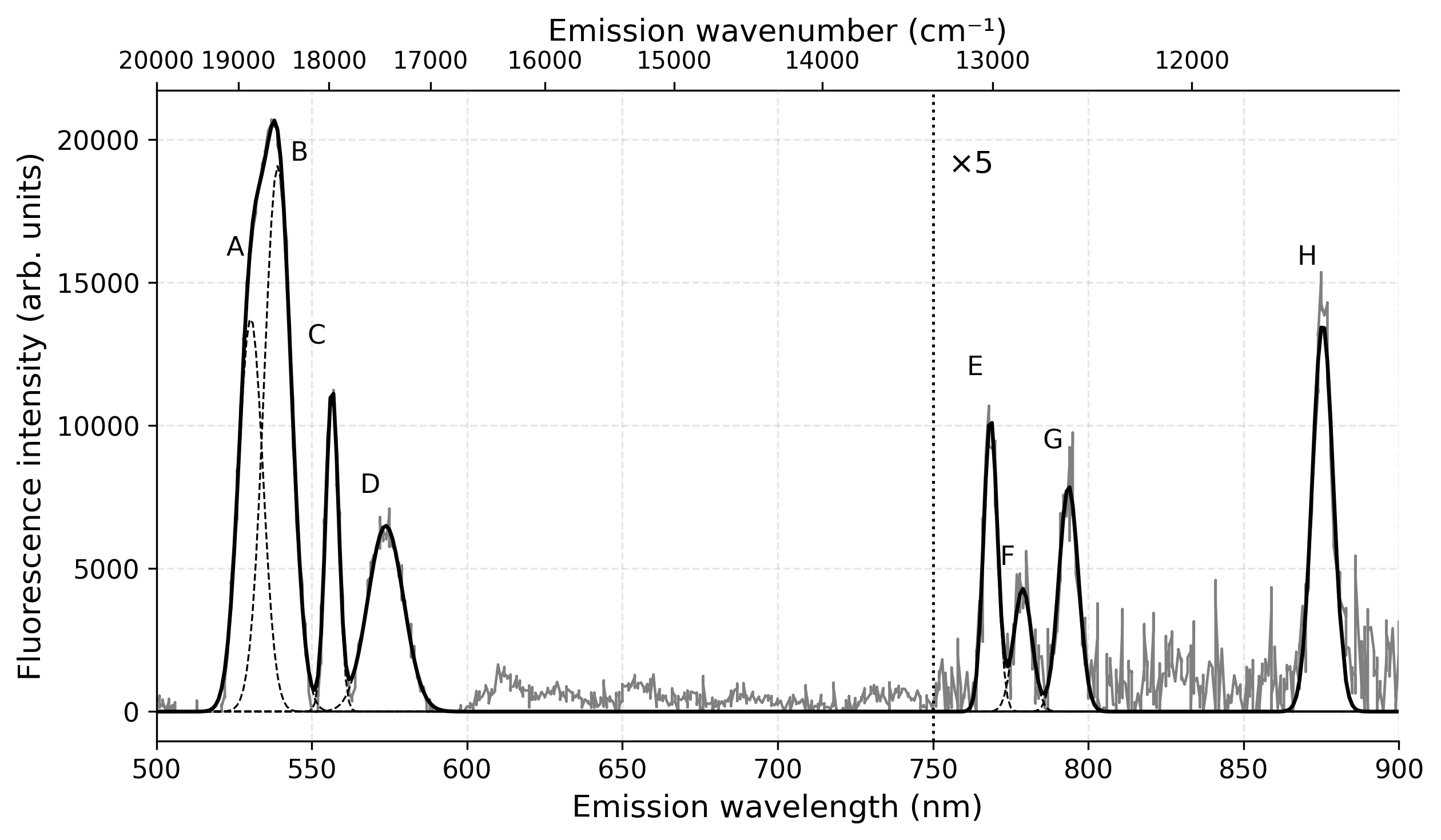}
    \caption{ Fluorescence spectrum of atomic barium in a neon cryogenic matrix at 6.8 K when excited with a third harmonic (355 nm) of a Nd:YAG laser with a repetition rate of 10 Hz. In gray: experimental data (baseline-subtracted); black solid: multi-Gaussian total fit; black dashed: individual fitted transitions. The region above 750 nm is magnified ×5. List of cascade fluorescence detected:  A - 6s9s $^3\text{S}_1$ to 6s6p $^1\text{P}_1$; B - 6s6d $^3\text{D}_1$ to 6s6p $^3\text{P}_0$; C - 6s6p $^1\text{P}_1$ to 6s$^2$ $^1\text{S}_0$; D - 6s6d $^3\text{D}_3$ to 6s6p $^3\text{P}_2$; E - 5d6p $^3\text{F}_2$ to 5d6s $^3\text{D}_1$; F - 5d6p $^3\text{F}_2$ to 5d6s $^3\text{D}_2$; G - 6s6p $^3\text{P}_1$ to 6s$^2$ $^1\text{S}_0$; H - 5d6s $^1\text{D}_2$ to 6s$^2$ $^1\text{S}_0$. }

    \label{3rd_harm}
\end{figure}

The work of Lebedev et al. \cite{Weis_BaHelium} demonstrate that a third harmonic of an Nd:YAG laser, with pulses at 28\,195 cm$^{-1}$ (355 nm) can excite a wide range of barium states by both radiative or non-radiative transitions in a solid helium matrix. Among these, long lived metastable state of barium, such as 5d6s $^1$D$_2$ and the 5d6s $^3$D$_j$ ($j = 1, 2, 3$) manifold, act as an intermediate reservoir, and when a second pulse excites them, they can reach excited state at energies over 30\,000 cm$^{-1}$ with respect to the energy of the ground state. This method leads to barium fluorescence emission through a radiative cascade, as from the highest excited state the atom can populate a metastable state before decaying into the ground state.
\\
We adapted this approach to detect barium fluorescence in a neon crystal (vis-NIR), enabling the first comparison up to date with the work of Lebedev et al. \cite{Weis_BaHelium} in solid helium.
\\
As illustrated in Figure \ref{3rd_harm} and summarized in Table \ref{Table_3rd_harm} we observe a broad set of barium fluorescence peaks when a third harmonic of an Nd:YAG pulsed laser is shined onto the crystal. The reported spectra is obtained by subtracting the fluorescence spectra of the CaF$_2$ substrate from the one obtained from the barium doped neon matrix on the substrate.
\begin{table}[htbp]
\centering
\caption{Barium emission transitions in a neon crystal at 6.8 K were observed following excitation with the third harmonic (355 nm) of a Nd:YAG laser. The spectral shifts are referenced against the corresponding free atomic barium lines, listed in the lower portion of the table. Transition H is an electric dipole forbidden line in the gas phase \cite{NIST_ASD}. Symbol $\ast$ = multiple transitions involved.}
\label{Table_3rd_harm}
\begin{tabular}{ccccc}
\toprule
\multicolumn{5}{c}{Observed LIF peaks} \\
Label & $\lambda$ (nm) & ${\nu}$ (cm$^{-1}$) & Shift $\delta$ (cm$^{-1}$) & FWHM $\Delta \nu$ (cm$^{-1}$) \\
\midrule
A & 530.3 & 18\,854 & $+12$ & 330 \\
B & 539.0 & 18\,551 & $+122$ & 354 \\
C & 556.5 & 17\,968 & $-93$ & 156 \\
D & 573.8 & 17\,427 & $+124$ & 411 \\
E & 768.5 & 13\,011 & $-18$ & 88 \\
F & 778.9 & 12\,838 & $-9$ & 113 \\
G & 793.7 & 12\,599 & $-35$ & 117 \\
H & 875.5 & 11\,422 & $+27$ & 102 \\
\midrule
\multicolumn{5}{c}{Corresponding free atom transitions} \\
Label & $\lambda$ (nm) & ${\nu}$ (cm$^{-1}$) & Free gas transition \\
\midrule
A & 530.6 & 18\,842 & 6s9s $^3$S$_1$ - 6s6p $^1\text{P}_1$ \\
B & 542.5 & 18\,429 & 6s6d $^3$D$_1$ - 6s6p $^3\text{P}_0$ \\
C & 553.5 & 18\,061 & 6s6p $^1\text{P}_1$ - 6s$^2 \; ^1\text{S}_0$ \\
D & $\ast$ 577.9 & 17\,303 & 6s6d $^3\text{D}_3$ - 6s6p $^3\text{P}_2$ \\
E & 767.2 & 13\,030 & 5d6p $^3\text{F}_2$ - 5d6s $^3\text{D}_1$ \\
F & 778.0 & 12\,848 & 5d6p $^3\text{F}_2$ - 5d6s $^3\text{D}_2$ \\
G & 791.1 & 12\,635 & 6s6p $^3\text{P}_1$ - 6s$^2 \; ^1\text{S}_0$ \\
H & 877.5 & 11\,395 & 5d6s $^1\text{D}_2$ - 6s$^2 \; ^1\text{S}_0$ \\
\bottomrule
\end{tabular}
\end{table}
We identified a total of eight fluorescence peaks. Comparing our data with the spectroscopic results of Lebedev et al. on barium in solid helium \cite{Weis_BaHelium}, four fluorescence lines match those previously reported.
\\
In general, the relatively narrow wavelength shifts with respect to the free barium atomic lines and the relatively small full width at half maximum (FWHM) allow us to reliably identify five electronic state transitions (C, E, F, G, H, see Figure~\ref{3rd_harm}), as shown in Table~\ref{Table_3rd_harm}. Among these, the largest observed shift is for peak C ($-93$ cm$^{-1}$), while the smallest is for peak F ($-9$ cm$^{-1}$). The magnitude of these shifts is smaller than those reported for barium isolated in rare gas matrices \cite{McCaffrey}.
\\
However, some emission features may not be fully resolved because of their relatively large FWHM. Peaks A and B are identified through a multi-Gaussian fit of the same broad feature, as shown in Figure \ref{3rd_harm}, and are identified with small shifts of $+12$ and $+122$ cm$^{-1}$, respectively, with respect to the one measured in free atomic Ba. 
Peak D, centered at 17\,427 cm$^{-1}$ (573.8 nm), might arise from a convolution of multiple energy levels. In this wavelength range, transitions such as 6s6d $^3\text{D}_{1,2,3}$ to 6s6p $^3\text{P}_1$ (17\,091, 17\,236, and 17\,303 cm$^{-1}$), or 6p$^2$ $^3\text{P}_2$ to 6s6p $^1\text{P}_1$ (17\,284 cm$^{-1}$), could all contribute to the observed line, explaining its large FWHM (411 cm$^{-1}$). The reported shift in Table \ref{Table_3rd_harm} is referenced to the 17\,303 cm$^{-1}$ transition. 
\\
The FWHM values can be directly compared with those obtained by Lebedev et al.~\cite{Weis_BaHelium} of barium in solid helium at 1.5 K. Peak C, corresponding to the 6s6p $^1\text{P}_1$ to 6s$^2$ $^1\text{S}_0$ transition, has a width of $\Delta\nu = 156\ \text{cm}^{-1}$, broader than the helium value ($\delta\nu = 30\ \text{cm}^{-1}$). However, peak E, F and G show the same width with respect to the same transitions observed in solid helium.
\\
Overall, both the spectral shifts and FWHM measured here are consistent with those reported in solid helium.


\subsection{Excitation with single and double laser}
\label{avesta}
In this section, our objective is to investigate the doped crystal using continuous-wave lasers, following the work of Li et al. \cite{Li_2023}. To this end, we irradiated 1 mm-wide spots on the crystal surface with Ti:Sa lasers delivering up to 100 mW, and measured excitation spectra by tuning the laser over its operational range of 14\,300–11\,100 cm$^{-1}$ (700–870 nm).

\subsubsection*{Single laser excitation} 
\begin{figure}[]
    \centering   
    \includegraphics[width=60mm]{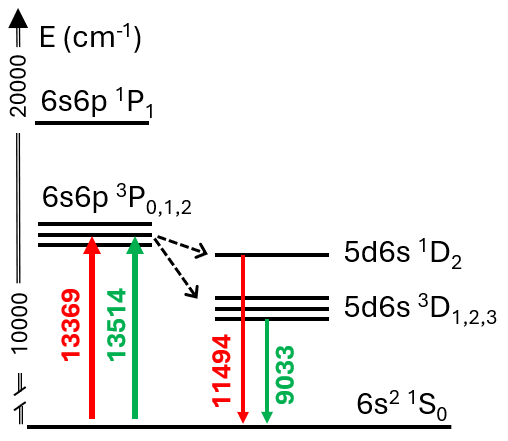}
    \caption{ Energy level diagram (not to scale) of the single laser pumping scheme of neutral barium atoms in a neon matrix. Both excitation lasers, at 13\,514 cm$^{-1}$ (749.8 nm, green) and 13 369 cm$^{-1}$ (740.5 nm, red), populate the 6s6p $^3$P$_j$ manifold. The system is relaxed to the states 5d6s $^3$D$_1$ and 5d6s $^1$D$_2$, which then decay to the ground state. These population transitions are detected respectively at 9\,033.4 cm$^{-1}$ (1107.1 nm, green) and 11\,494.2 cm$^{-1}$ (870.0 nm, red). } 
    \label{singlelaser_scheme}
\end{figure}
\begin{figure}[]
    \centering   
    \includegraphics[width=85mm]{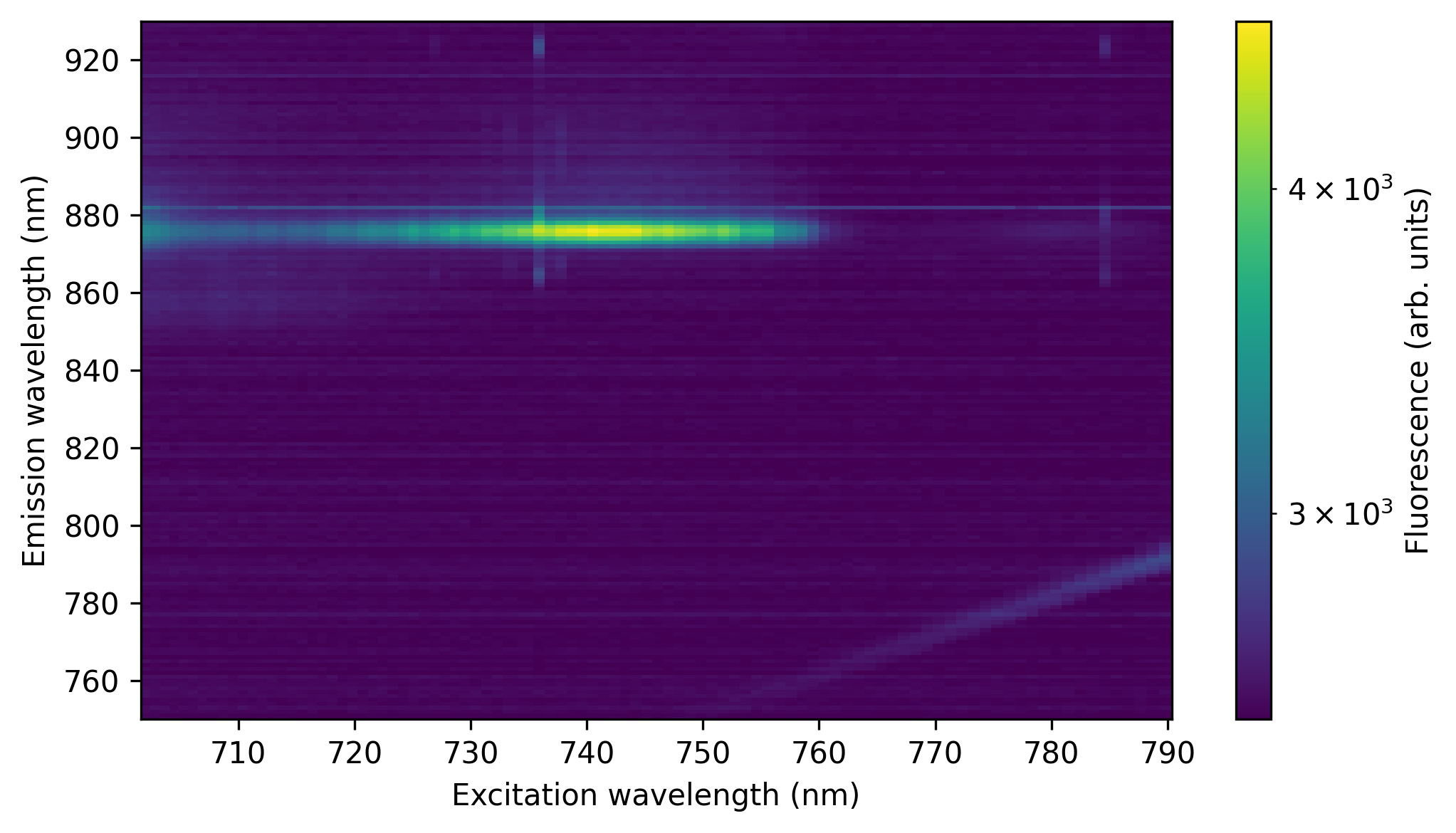}
    \caption{ 2D excitation–emission spectrum of barium in a neon crystal at 6.8 K when shined with a single laser at 35 mW on a 1 mm wide spot. A strong emission band is observed, with a maximum intensity at 11\,415 cm$^{-1}$ (876.0 nm), corresponding to population of the 5d6s $^1$D$_2$ state ($+20$ cm$^{-1}$ shift), when the laser is tuned at 13\,504.3 cm$^{-1}$ (740.5 nm). A long blue tail is visible until the reach of our laser at 700 nm of excitation. The residual signal of the excitation laser appears at longer wavelengths due to the reduced efficiency of the optical filtering.  } 
    \label{nIR_1laser}
\end{figure}
\begin{figure}[]
    \centering   
    \includegraphics[width=85mm]{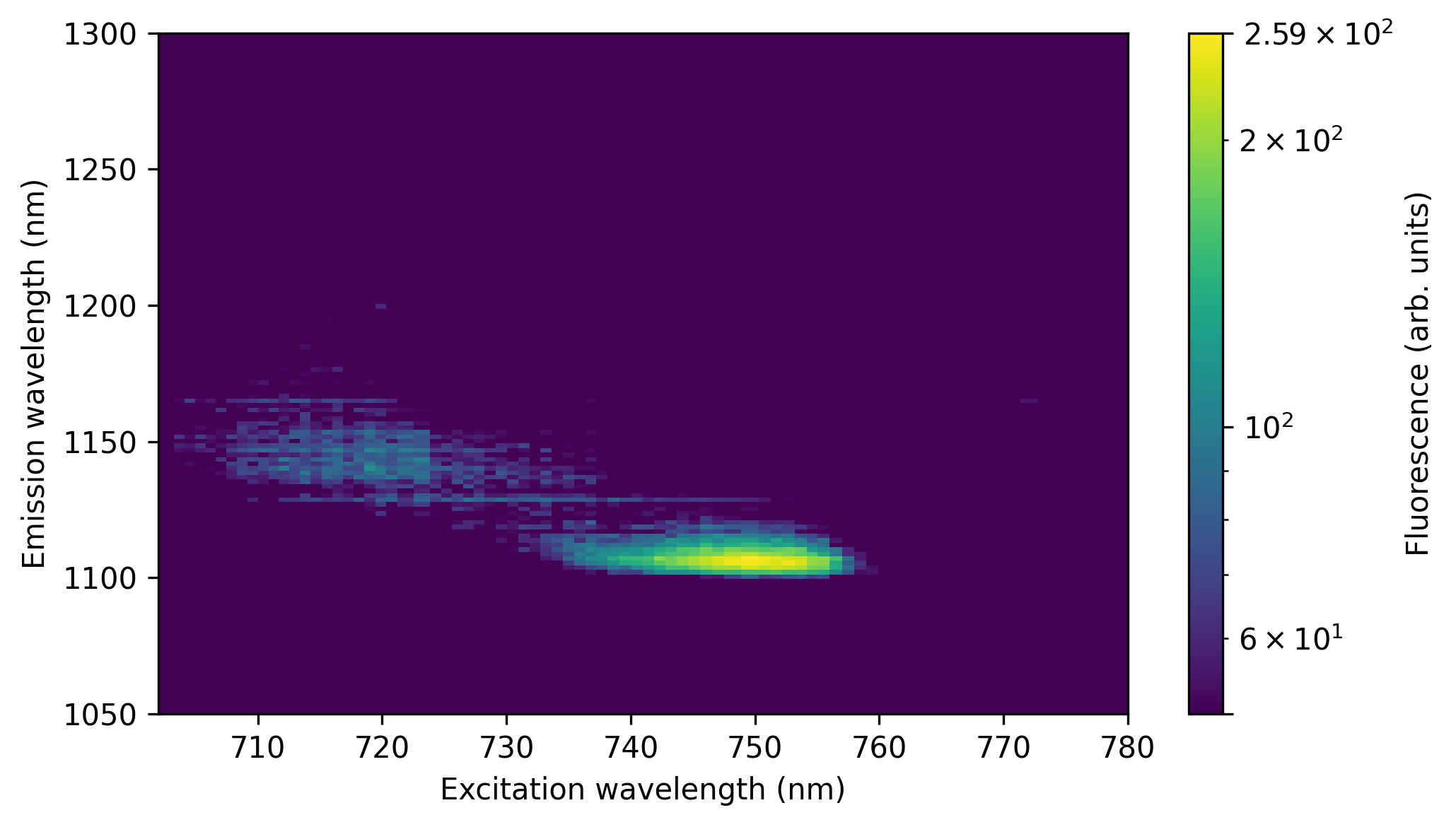}
    \caption{ 2D excitation–emission spectrum of barium in a neon crystal at 6.8 K when shined with a single laser at 35 mW on a 1 mm wide spot. An emission band is observed around 9033.4 cm$^{-1}$ (1107.1 nm), corresponding to population of the 5d6s $^3$D$_1$ state, with maximum intensity under excitation at 13\,514 cm$^{-1}$  (749.8 nm) and long blue tail of approximately 40 nm, similar to the one observed previously in Figure \ref{nIR_1laser}. } 
    \label{IR_1laser}
\end{figure}
\begin{figure}[]
    \centering   
    \includegraphics[width=90mm]{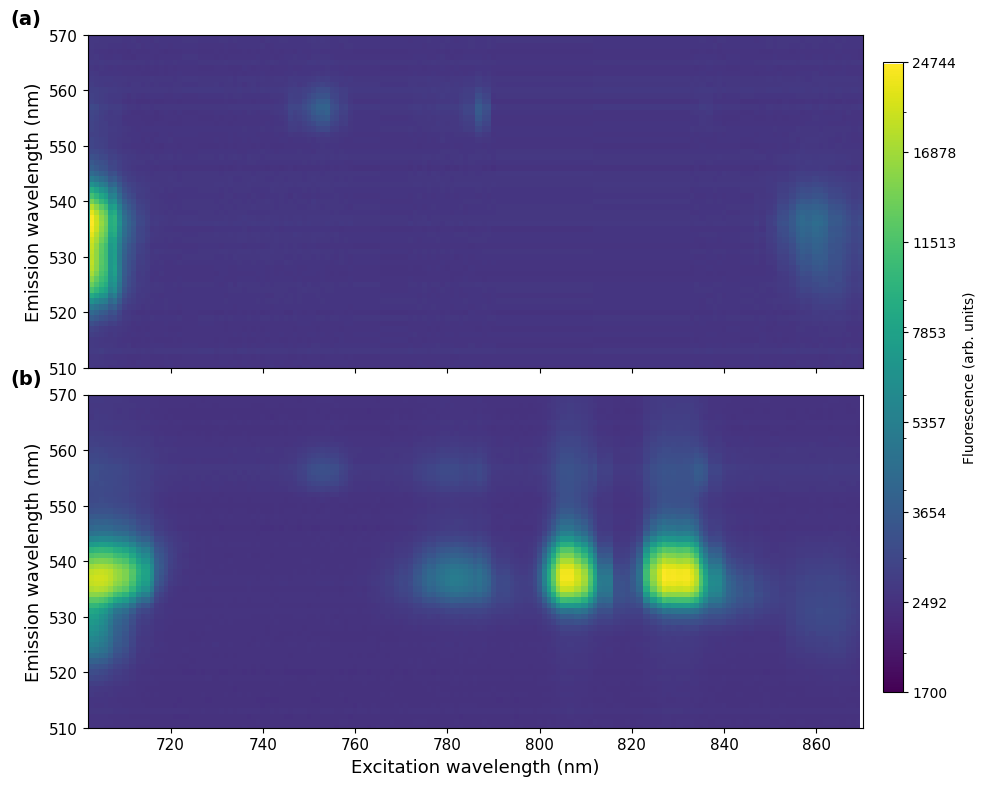}
    \caption{  2D excitation–emission spectra of barium atoms embedded in a neon crystal at 6.8 K, showing fluorescence in the green region. (a) Excitation with a single narrow linewidth laser at low energy allows population transfer from the first excited manifold to higher-lying states, whose fluorescence cascade produces green light comparable to that observed under third-harmonic excitation. (b) When a second identical laser is spatially overlapped with the first and tuned to 748 nm, which populates the 5d6s $^3$D$_1$ state, additional population transfer is observed. New fluorescence pathways emerge, revealing alternative excitation channels to higher-lying states } 
    \label{green_doubleplot}
\end{figure}

Radiative fluorescence to the ground state  6s$^2 \, ^1\text{S}_0$ is observed from the metastable states 5d6s $^3$D$_1$ at 9\,033.4 cm$^{-1}$ (1107.1 nm) and 5d6s\;$^1$D$_2$ at 11\,494.2 cm$^{-1}$ (870.0 nm) after excitation around 13\,514 cm$^{-1}$ (749.8 nm) and 13\,504 cm$^{-1}$ (740.5 nm) respectively, as shown in Figure \ref{singlelaser_scheme}. Those transitions are forbidden in the free gas phase and, as shown in Figures \ref{nIR_1laser} and \ref{IR_1laser}, both states are populated by broad excitation bands with blue tails that extend up to 700 cm$^{-1}$ (40 nm) from the center of the two peaks. 
\\
These excitation wavelengths are far detuned from direct transitions and likely populate the 6s6p $^3$P$_j$ manifold through matrix-induced relaxation, as shown in Figure \ref{singlelaser_scheme}. Those states can decay to the ground state through radiative cascade involving the 5d6s $^3$D$_i$ (i = 1,2,3) and $^1$D$_2$ states. Direct fluorescence from the $^3$P$_j$ states is not detected, as it might be too close to the excitation light which was properly filtered out. Stoke shifts are usually big enough to directly detect emission light, but in neon such transitions have not been detected in the spectral region of collection. In the matrix environment, the local perturbation breaks the atomic symmetries that govern the selection rules in the gas phase \cite{moroshkin2008atomic}. As a result, transitions that are forbidden for free barium become weakly allowed. In our case, the excitation laser seems to directly address the 6s6p $^3$P$_2$ level at 13\,514.7 cm$^{-1}$ (740.0 nm), a strongly suppressed transition in the free atom, allowing the population transfer to the 6s6p $^3$P$_2$ state and the subsequent relaxation into the 5d6s $^3$D$_i$ manifold or 5d6s $^1$D$_2$ state.
\\
As stated in precedent spectroscopy studies of Ba atoms in RG gas matrices \cite{McCaffrey}, the line broadening of excitation is larger with respect to the fluorescence emission. In this case, the effect is likely enhanced by the indirect excitation discussed above.
\\
Green fluorescence around 18000 cm$^{-1}$ corresponding to the 6s6p $^1$P$_1$ state has been detected, as shown in Figure \ref{green_doubleplot} (a). 
\\
The photon energies provided by our excitation laser are insufficient to directly excite the 6s6p $^1$P$_1$ state. Our hypothesis, further investigated in the following sections, is that since the photon energy of the Ti:Sa laser is insufficient to reach the 6s6p,$^1$P$_1$ state directly, the observed green fluorescence likely arises from multi step population of higher manifolds even under single frequency continuous excitation. This mechanism is analogous to the excitation scheme achieved with the 3rd harmonic pulsed laser, as discussed in Section \ref{Barium_third}.

\subsubsection*{Double laser excitation scheme and results}

To test the feasibility of pumping schemes with continuous lasers, we performed further measurements using a combination of two identical Ti:Sa laser.
\\
Since our strongest source of fluorescence signal is the green one, we first focused on it. The first laser is fixed at the peak of population transfer in the 5d6s $^3$D$_1$ state (9\,033.4 cm$^{-1}$), while we scanned with the second excitation laser over the same frequency region as before (14\,300 - 11\,100 cm$^{-1}$), overlapping the two laser interaction regions in the crystal at a power of 40 mW. 
\\
The resulting spectrum, shown in Figure \ref{green_doubleplot} (b), reveals an increase in the fluorescence intensity of the previously detected signal, as well as the appearance of new excitation combinations, resulting in high intensity fluorescence of the 6s6p $^1$P$_1$ state. 
\\
The 6s6p $^3$P$_1$ state exhibits two dominant fluorescence bands centered at 18\,622 (557 nm) and 17\,967 (538 nm) cm$^{-1}$. Similar features have been extensively studied in Ar, Kr, and Xe matrices \cite{McCaffrey}, where they are attributed to the influence of lattice site occupancy on barium emission properties. Further investigation of the process of site selection of Barium in a neon crystal is out of the scope of our project and has already been discussed for BaF embedded in a neon matrix \cite{lambo2023calculationlocalenvironmentbarium}.
\begin{figure}[]
    \centering   
    \includegraphics[width=60mm]{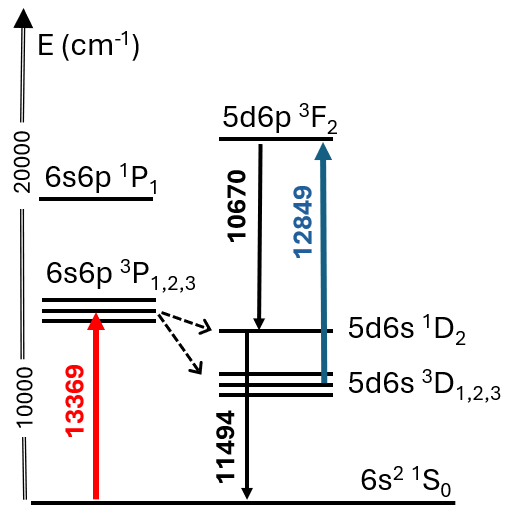}
    \caption{ Energy level diagram (not to scale) of the proposed double laser pumping scheme for barium in a neon matrix. The first laser (red arrow) at 13\,369 cm$^{-1}$ (749 nm) indirectly populates the 5d6s $^3$D$_2$ state. From this level, a second laser (blue arrow) tuned at 12\,849 cm$^{-1}$ (780 nm) excites the atom to the higher 5d6p $^3$F$_2$ state. According to the cascade dynamics reported in previous works \cite{Weis_BaHelium}, this state decays to the 5d6s $^1$D$_2$ level with emission at 10\,670 cm$^{-1}$ (937 nm), followed by a second fluorescence decay to the ground state 6s$^2$ $^1$S$_0$ at 11\,394 cm$^{-1}$ (877 nm). The black arrows represent spontaneous emission pathways, while red arrows denote laser driven excitations. Dashed lines indicate indirect population routes from the 6s6p $^3$P$_j$ manifold. } 
    \label{940_pumping scheme}
\end{figure}
\\
We observe a high intensity fluorescence emission when the excitation laser is tuned to approximately 14\,180 cm$^{-1}$ (705 nm) which is probably populating the 5d6s $^1$D$_2$ state as shown in Figure 5, as well as two dominant emission when a second laser excites the sample at around 12\,270 cm$^{-1}$ (815 nm) and 12\,060 cm$^{-1}$ (830 nm). We note that the three excitations are close to BaF energy levels reported in previous works \cite{Li_2023}. The observed emission is originating from different population pathways involving long living intermediate states, as shown in the following sections. 
\\
To further test the feasibility of the double–laser excitation, we designed a pumping scheme path that populates 5d6p $^3$F$_2$ at 22\,065 cm$^{-1}$ (453 nm) only if we strongly populate the manifold 6s6d $^3\text{D}_j$, as shown in Figure \ref{940_pumping scheme}. Although the excitation bands are $\approx$1000 cm$^{-1}$ wide, our measurements demonstrate that this scheme works exclusively via indirect population of the first manifold, specifically through the 5d6s $^3$D$_2$ state. A second laser, set at 12\,849 cm$^{-1}$, can then couple this state and populate the 5d6p $^3$F$_2$ state laying at \num{22065} cm$^{-1}$ (453 nm).
This scheme relays on the evidence shown earlier that excitation near 13\,369 cm$^{-1}$ (748 nm) populates the 5d6s $^3$D$_1$ and hypothetically the other states of the same 5d6s $^3$D$_j$ manifold.
\begin{figure}[]
    \centering   
    \includegraphics[width=90mm]{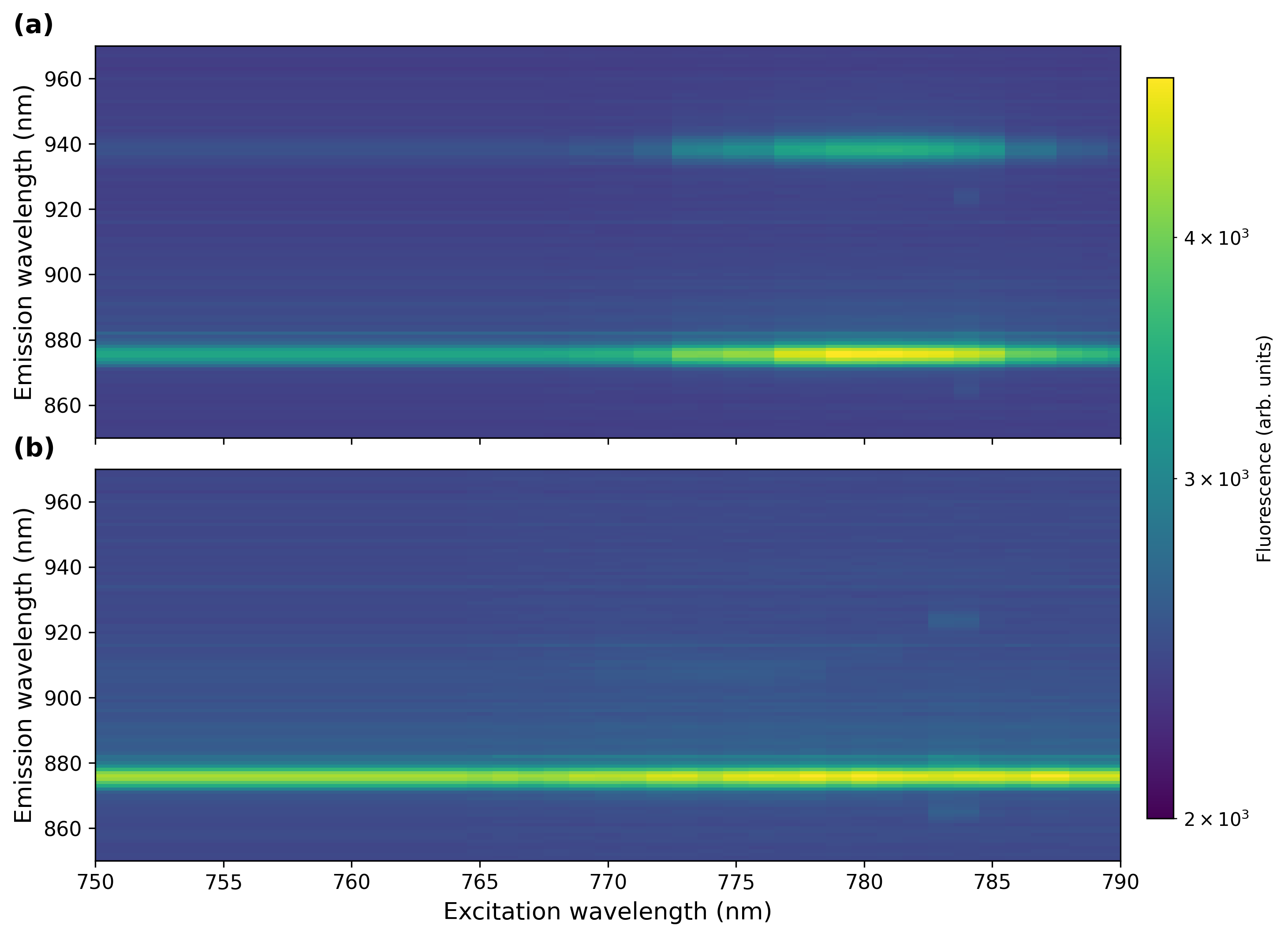}
    \caption{ 2D excitation–emission spectra of the double-laser test probing the 5d6p;$^3$F$_2$ fluorescence cascade of barium in a neon matrix.
(a) Two spatially overlapped continuous-wave beams (40 mW each) illuminate the crystal: the first, fixed at 13\,369 cm$^{-1}$ (749 nm), populates the first excited-state manifold (5d6s $^3$D$_2$), while the second is scanned near 12\,849 cm$^{-1}$ (780 nm) to drive the 5d6s $^3$D$_2$ to 5d6p $^3$F$_2$ transition. A distinct fluorescence band at 10\,670 cm$^{-1}$ (937 nm) appears, together with an enhanced 11\,395 cm$^{-1}$ (877 nm) band, consistent with the cascade 5d6p $^3$F$_2$ to 5d6s $^1$D$_2$ (937 nm) followed by 5d6s $^1$D$_2$ to the ground state (870 nm).
(b) When the first laser is tuned to 740 nm, the 937 nm band disappears while the 877 nm baseline remains, confirming that the previous excitation scheme selectively populates the 5d6s;$^3$D$_2$ state, which is required to reach 5d6p $^3$F$_2$.} 
    \label{940_results_comparision}
\end{figure}
\\
The expected result of the pumping scheme is a cascade fluorescence, starting from the level 5d6p $^3$F$_2$ and decaying to $^1$D$_2$ at 10\,670 cm$^{-1}$ (937 nm), followed by an emission to the ground state at 11\,395 cm$^{-1}$ (878 nm), as indicated by the straight black lines in Figure \ref{940_pumping scheme}.
\\
Experimentally, the first laser is fixed at 13\,369 cm$^{-1}$, while the second is scanned between 13\,300 and 12\,700 cm$^{-1}$. The result, shown in Figure \ref{940_results_comparision} (a), reveals expected LIF emissions at 10\,640 cm$^{-1}$ (939 nm) along with an enhanced line at approximately 11395 cm$^{-1}$ (878 nm), but only when the second laser is tuned around 12\,805 cm$^{-1}$ (781 nm), close to the expected wavelength, shifted by just 44 cm$^{-1}$. This observed fluorescence confirms the predicted cascade behavior from 5d6p $^3$F$_2$ to 5d6s $^1$D$_2$ and then to the ground state. 
\\
To further test the scheme, if we tune the first laser to 13\,504 cm$^{-1}$ (740 nm), a known population maximum of the 5d6s $^1$D$_2$ state, no excitation pathway to the 5d6p $^3$F$_2$ level is observed, as shown in Figure \ref{940_results_comparision} (b), probing our selective pumping scheme functioning. Moreover, both lasers have been applied to the crystal at higher intensities, individually and simultaneously, as saturation is usually never reached in a matrix isolation environment and additional atom emissions might arise. However, the obtained spectra exhibit the same characteristic as those reported here.


\subsection{Lifetime of first excited state 5d6s $^3$D$_1$}
\begin{figure}[]
    \centering   
    \includegraphics[width=90mm]{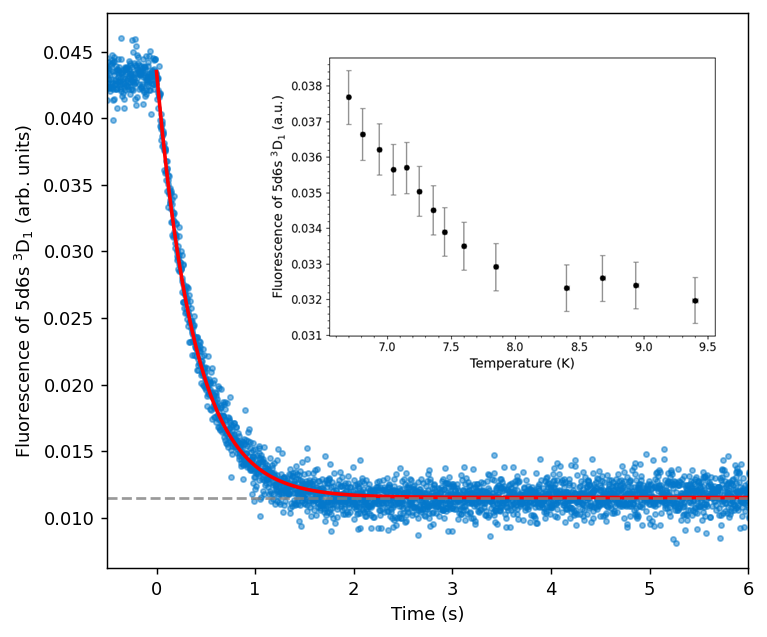}
    \caption{ Fluorescence decay of the 5d6s $^3$D$_1$ state of barium embedded in solid neon at 6.8 K. The red curve shows an exponential fit, yielding a lifetime determination of 0.39 $\pm$ 0.02 s. The gray dashed line represents the background level of fluorescence light due to imperfect extinction of the excitation light when the EOM is off. Inset: Fluorescence intensity dependence on the crystal temperature. At temperature higher than 8 K, intensity in the fluorescence shows a plateau.  
} 
    \label{lifetime_1stExc}
\end{figure}

As shown in the previous sections, and schematically in Figure \ref{singlelaser_scheme}, a single laser allows us to populate the 5d6s $^3$D$_j$ manifold, similar to the earlier work of Lebedev et al. \cite{Weis_BaHelium}, where the 5d6s $^3$D$_1$ state of barium in a helium matrix at 1.5 K is populated using the pulsed second harmonic of a Nd:YAG laser at approximately $18\,800$ cm$^{-1}$ .
\\
In our experiments with barium in a solid neon at 6.8 K, this state is populated using continuous wave laser light tuned to 13\,514 cm$^{-1}$, enabling a direct comparison with the lifetime measurement performed in solid helium.
\\
The lifetime measurement is carried out using an Electro Optic Modulator (EOM), driven by a pulse generator. This configuration allowed us to modulate the excitation beam intensity, switching the laser on and off in 200 ns.
\\
The decay of the state population from 5d6s $^3$D$_1$ to the ground state is shown in Figure \ref{lifetime_1stExc}, as well as the fit to compute the measurement of the lifetime ($\tau$) following the model: $I = I_0 e^{-t/\tau} + C$, where $I_0$ is the initial intensity and $C$ is the background intensity due to the non-zero extinction of the EOM. This background level is shown in Figure \ref{lifetime_1stExc} as a gray dashed line. 
\\
The resulting lifetime of the 5d6s $^3$D$_1$ state of barium in the neon matrix is found to be $0.39 \pm 0.02$ s.
For a comparison, in solid helium, a longer lifetime of $2.72 \pm 0.04$ s is observed at a temperature of 1.5 K \cite{Weis_BaHelium}. 
\\
Lifetime measurements have been conducted at different temperatures, as, ranging from 6.8 to 8 K, due to temperature limitations of the setup. 
A measurement of the LIF intensity depending on the substrate temperature is shown in the inset of Figure \ref{lifetime_1stExc}. At temperatures higher than 8 K a plateau is observed. This is an unknown phenomena, that might be due to the sublimation of the neon crystal starting at approximately 8 K in a vacuum of $10^{-7}$ mBar.
\begin{figure}[]
    \centering   
    \includegraphics[width=95mm]{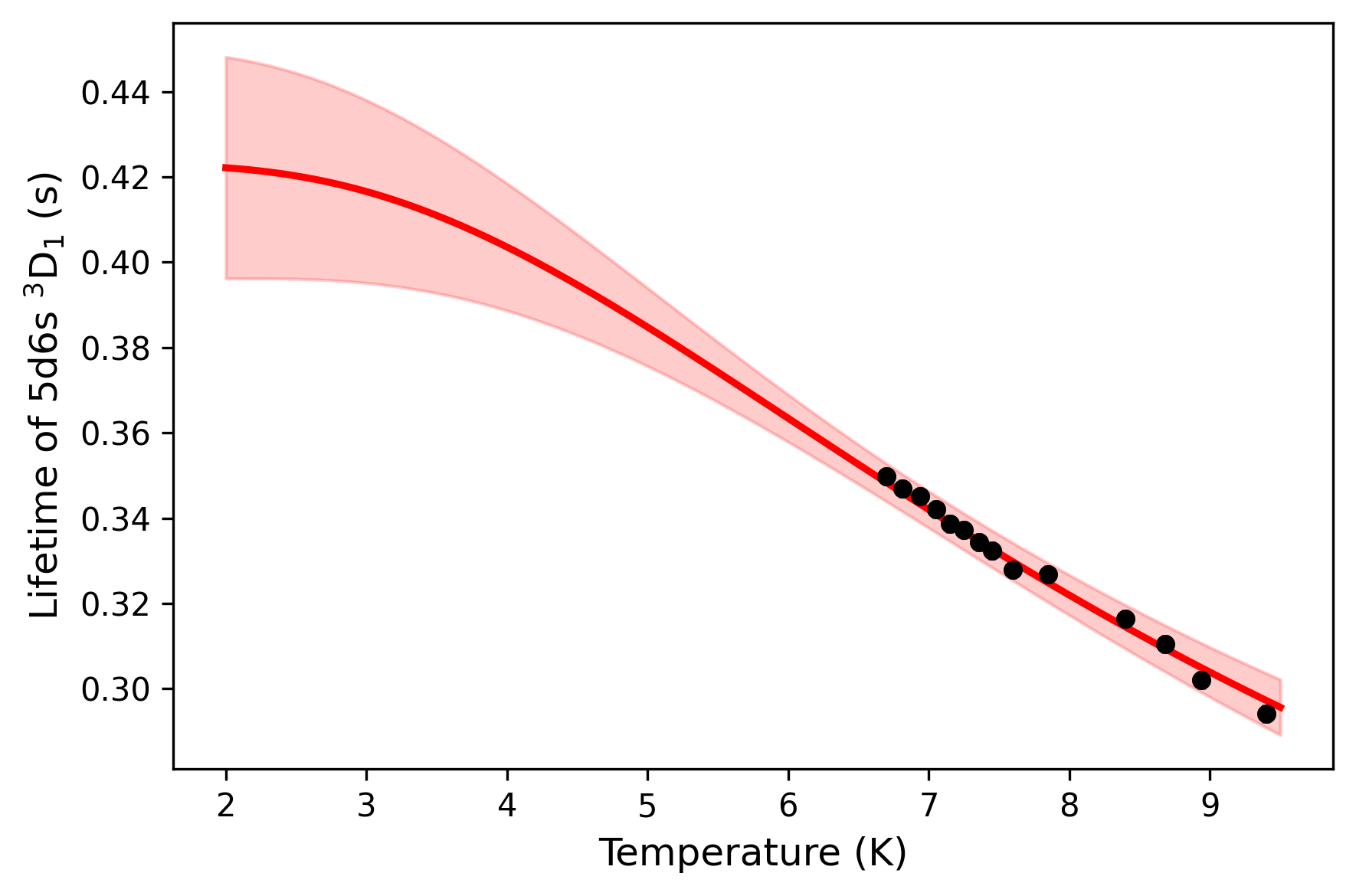}
    \caption{ Temperature dependence of the lifetime of the 5d6s;$^3$D$_1$ state of barium embedded in solid neon. In red a fit of Eq. \ref{fit_r_nr}, the rate equation governing the radiative and non-radiative decay of an atom embedded in a cryogenic crystal lattice. From such a fit, a lifetime of 0.42 $\pm$ 0.03 s is expected at 2 K. The shaded red band represents the one $\sigma$ confidence interval of the fit. } 
    \label{tauVST}
\end{figure}
\\ 
A rate equation model has been used to fit the obtained lifetime and extrapolate the possible lifetime values at lower temperatures. It considers both radiative ($\tau_r$) and non-radiative ($\tau_{nr}$) decay channels, described by the two terms\cite{fit_description_PL, torVergata_fitArrhenius}: 
\begin{equation}
\label{fit_r_nr}
    \frac{1}{\tau(T)} \;=\; \frac{1}{\tau_r} \;+\; \frac{1}{\tau_{nr}} \, \exp\!\left(-\frac{E_a}{k_B T}\right) \, .
\end{equation}
The fit results are shown in Table \ref{tab:rate_model_fit}, where $E_a$ is the activation energy. The lifetime is expected to increase to approximately 0.42 $\pm$ 0.03 s at 2 K. T
The model captures the expected low temperature plateau corresponding to a regime where the dopant non radiative decay channels are suppressed. Moreover, in the precedent work of BaF in neon\cite{Li_2023}, a similar model has been used, with a resulting activation energy value of $E_{a (BaF:Ne)}  = hc \times 16.2(3) \; \text{cm$^{-1}$} = (2.01 \pm 0.04)\times 10^{-3} \; \text{eV}$, consistent with the one obtained in this work, shown in shown in Table \ref{tab:rate_model_fit}.
Future work and new models might take into account additional mechanisms inherent to matrix environments, such as site dependent trapping, crystal inhomogeneity and local stress of laser excitation light. 
\begin{table}[] 
\centering
\caption{Fit parameters from the rate equation model describing the temperature dependent lifetime of the 5d6s $^3$D$_1$ state of barium in solid neon.}
\begin{tabular}{lcc}
\hline
\textbf{Parameter} & \textbf{Value} \\
\hline
$\tau_r$ & $0.42 \pm 0.02$ s \\
 $\tau_{nr}$ & $0.16 \pm 0.01  $ s \\
$E_a$ & $(1.4 \pm 0.3) \times 10^{-3}$ eV \\
\hline
\end{tabular}
\label{tab:rate_model_fit}
\end{table}

\section{Conclusions}

For the first time in literature, we observed barium atoms embedded in a neon matrix through a combination of cascade fluorescence and selective excitation using a third harmonic 1064 nm pulsed laser and single or double continuous wave lasers, exploiting long living metastable state of barium within the matrix.
\\
The observed fluorescence spectra display small shifts relative to the gas phase transitions and moderate inhomogeneous broadening, indicating the weak perturbation exerted by the neon matrix. 
The main 6s6p $^1$P$_1$ to 6s$^2$ $^1$S$_0$ transition was observed at 556.5 nm, with a shift with respect to the free gas phase of $-93$ cm$^{-1}$ and a linewidth of 156 cm$^{-1}$. 
Overall, all identified transitions have shown relatively small matrix shifts and widths with respect to heavier RG matrices (Ar, Kr, Xe). The 5d6p $^3$F$_2$ to 5d6p $^3$P$_j$ transitions show shifts comparable to those in helium, confirming neon’s low perturbation nature.
\\
Using a single tunable laser between 700 and 900 nm, we identified indirect excitation pathways populating the 5d6s $^3$D$_1$ and $^1$D$_2$ metastable states via matrix induced symmetry breaking.
With two simultaneously applied lasers, we demonstrated an excitation pathway from the 5d6s $^3$D$_2$ state to 5d6p $^3$F$_2$, followed by a cascade fluorescence sequence 5d6p $^3$F$_2$ to 5d6s $^1$D$_2$ to 6s$^2$ $^1$S$_0$, confirming the feasibility of selective excitation within the matrix.
\\
This spectroscopy work on neutral barium in solid neon provides a crucial benchmark for future BaF spectroscopy within the same cryogenic matrix. Since the two species share similar spectral regions, and barium is produced during the production of BaF, a detailed understanding of Ba optical transitions, line widths, and fluorescence dynamics is essential to correctly identify and separate atomic and molecular contributions in mixed Ba/BaF matrices.
\\
Furthermore, we measured for the first time the lifetime of the 5d6s $^3$D$1$ state of barium in solid neon, finding a value of 0.39 $\pm$ 0.02 s, with an expected increase of 10$\%$ at 2 K. This long lived state confirms neon as a promising host for matrix isolation measurements with low perturbation from the matrix. In addition, solid neon offers a stable crystalline structure that allows long data acquisition times without signal degradation or sample loss \cite{Neon_matrix}.
\\
Our lifetime measurement serves as a reference point for future studies aiming to explore similar systems in parahydrogen matrices. Previous work by Lancaster et al. \cite{Lancaster_2021} found no detectable fluorescence for rubidium in parahydrogen, whereas clear emission was observed for boron \cite{Tam2000_boron_pH} and other molecules \cite{Huang2016_molecules_pH}. A similar investigation for Ba and BaF will be crucial to establish parahydrogen as a next-generation host for matrix-isolated spectroscopy and precision measurements of fundamental properties.
\begin{acknowledgments}
We express our sincere gratitude to A. C. Vutha for contributing to insightful discussions that greatly contributed to this work. 
\\

We also thank E. Berto, F. Calaon, M. Tessaro, M. Rebeschini, and M. Zago for their invaluable technical support.
\end{acknowledgments}

\section*{AUTHOR DECLARATIONS}
\subsection*{Conflict of Interest}
The authors have no conflicts to disclose.

\subsection*{Author contributions}
\textbf{Alessandro Lippi:} Conceptualization (equal); Data curation (lead); Formal analysis (equal); Investigation (equal); Methodology (equal); Software (lead); Visualization (lead); Writing – original draft (lead); Writing – review \& editing (equal).
\\
\textbf{Giovanni Carugno:} Conceptualization (equal); Funding acquisition (lead); Methodology (equal); Project administration (lead); Resources (equal); Supervision (lead); Writing – review \& editing (equal).\\
\textbf{Roberto Calabrese:} Funding acquisition (equal); Resources (equal).\\
\textbf{Federico Chiossi:} Validation (equal); Writing – review \& editing (equal).\\
\textbf{Marco Guarise:} Formal analysis (equal); Funding acquisition (equal); Resources (equal); Supervision (equal); Validation (equal); Writing – review \& editing (equal).\\
\textbf{Madiha M. Makhdoom:} Investigation (equal); Writing – review \& editing (equal).\\
\textbf{Giuseppe Messineo:} Conceptualization (equal); Investigation (equal); Validation (equal); Writing – review \& editing (equal).\\
\textbf{Jacopo Pazzini:} Funding acquisition (equal); Resources (equal); Writing – review \& editing (equal).\\

\subsection*{Founding}
This research was funded by the PNRR MUR project PE0000023 – NQSTI, financed by the European Union—Next Generation EU. This work was also supported in part by the Italian MUR “Departments of Excellence grant 2023–2027—Quantum Frontiers”, by the “Progetti di Ricerca di Rilevante Interesse Nazionale” (PRIN grant 20227F5W4N, PRIN 2022), and by INFN CSNV within the framework of DOCET experiment.

\section*{Data availability}
The data supporting the findings of this study are available
from the corresponding author upon reasonable request.



\nocite{*}
\bibliography{bibligraphy}

\end{document}